\documentclass[twocolumn,twoside,slac_two]{revtex4}
\usepackage{graphicx}
\usepackage{fancyhdr}
\newcommand{\filter}[1]{{\sc \lowercase{#1}} filter}
\newcommand{\lowe}{{\sc le}}
\newcommand{\highe}{{\sc he}}

\begin{document}

\title{Extending the Galactic Cosmic Ray electron + positron spectrum measured by the Fermi LAT}

%

\author{M. Pesce Rollins on behalf of the Fermi-LAT Collaboration}
\affiliation{INFN-Sezione di Pisa, I-56127 Pisa, Italia}
%

\begin{abstract}
Launched on the 11th of June 2008, the Fermi Large Area Telescope (LAT) has made
several outstanding scientific contributions to the high energy astrophysics 
community. One of these contributions was the high statistics measurement 
of the Galactic Cosmic Ray (GCR) electron + positron spectrum from 20 GeV 
to 1 TeV. The Fermi satellite is in a nearly circular orbit with an 
inclination of 25.6 degrees at an altitude of 565 km. Given this 
orbit it is possible to measure the GCR electrons + positrons down to 
roughly 5 GeV. 
However, this lower limit in energy is highly dependent on the orbital 
position of the LAT in geomagnetic coordinates due to the rigidity cutoff. 
In order to measure the spectrum down to these energies it is necessary 
to sample the population of electrons + positrons in several different 
geomagnetic positions. In this poster we present the analysis performed 
to extend the lower limit in energy of the GCR electron + positron spectrum 
measured by the Fermi LAT.  
\end{abstract}

\maketitle

\thispagestyle{fancy}

\section{Introduction}
The electron component of the Galactic Cosmic Ray (GCR) radiation is widely
recognized as a unique probe to address a number of significant questions
concerning the origin of cosmic rays and their propagation in our galaxy
(see \cite{crereview} for a review). For energies above a few GeV the dominant
energy loss mechanisms for electrons and positrons are the synchrotron
and inverse compton processes. These processes have an energy loss rate whose 
magnitude increases with the square of the electron energy and as a
consequence these electrons have a radiative lifetime inversely proportional
to their energy~\cite{HEAT}. Because of this relatively short radiative 
lifetime, the 
GCR electrons cannot travel intergalactic distances through the cosmic 
microwave background and thus excludes an extragalatic contribution to the 
measured spectra. After the first six months of nominal science data taking,
the Fermi LAT was able to make the first systematics limited measurement of
the GCR $e^-+e^+$ (referred to as electrons for the rest of this work) 
spectrum from 20 GeV to 1 TeV~\cite{PRL}. This measurement 
together with the recently published results from experiments such as 
PAMELA~\cite{pamela}, ATIC~\cite{atic} and HESS~\cite{hess} has set the stage 
for new and exciting GCR science. To help make constraints on possible 
propagation models it is important to extend the measurement of the GCR 
electrons to lower energies.  This task is the main topic of this work and 
the analysis used to extend the GCR electron spectrum measured by the Fermi 
LAT will be described in detail in this paper.

\section{Instrument description}
The Large Area Telescope, the main instrument onboard the Fermi observatory,
is a pair-conversion telescope composed of:
\begin{itemize}
\item[(i)] a tracker-converter, including 18 Silicon Strip Detector (SSD) $x-y$
  tracking planes, interleaved with tungsten conversion foils;
\item[(ii)] a CsI imaging calorimeter;
\item[(iii)] a segmented Anti-Coincidence Detector (ACD) surrounding the
  tracker subsystem.
\end{itemize}
Though it essentially follows in the footsteps of the previously flown
pair-conversion telescopes (particularly EGRET \cite{egret}) in its basic
design, the LAT is equipped with the state of the art in high-energy physics
detector technology and features many design improvements with respect
to its predecessor, resulting in a significant leap in sensitivity
(for a detailed description of the detector see \cite{LAT}).

From the standpoint of the electron detection, the tracker section effectively
acts as a finely segmented pre-shower (1.5 radiation lengths on axis) which
allows to image the initial part of the electromagnetic showers, providing
a crucial handle for the background rejection. The hodoscopic configuration
of the calorimeter (8.6 radiation lengths on axis) provides a full 
3-dimensional reconstruction of the shower
shape which is essential both for the energy reconstruction and for the 
discrimination between electromagnetic and hadronic
showers. The ACD also provides a relevant contribution to the background
rejection.

\subsection{Trigger and filter}
Two of the main design improvement of the LAT are a flexible trigger
logic and onboard event filtering.
The chosen detector technology allows a minimum dead time per event
of about 26 $\mu$s (essentially due to instrument readout). 
Therefore the LAT can trigger at the passage of (almost) every 
particle, including cosmic rays (only events that clearly behave as MIPs 
or heavy ions are hardware-prescaled to avoid the induction 
of unnecessary deadtime).
This is a revolution with respect to the previous generation 
of pair conversion  telescopes, in which 
the instrumental deadtime forced to apply a harsh 
photon selection at the level of the hardware trigger. 
The LAT trigger rate is of the order of few kHz but due to telemetry 
bandwidth limitations an onboard event selection is applied and roughly 
half of the events are actually written on spacecraft memory. 
This is accomplished by several software configurable onboard filters. 
The rate of events downlinked to Earth is of the order of few hundred Hz, 
still mostly charged particles (i. e. protons, 
electrons, GCR metals). The refined analysis, both for the photon  
(cfr.\cite{LAT}) and CR electron science is performed on the ground.
For more details about the on-orbit rates, see \cite{Eduardo}.
\begin{figure*}[t]
\centering
\includegraphics[width=130mm]{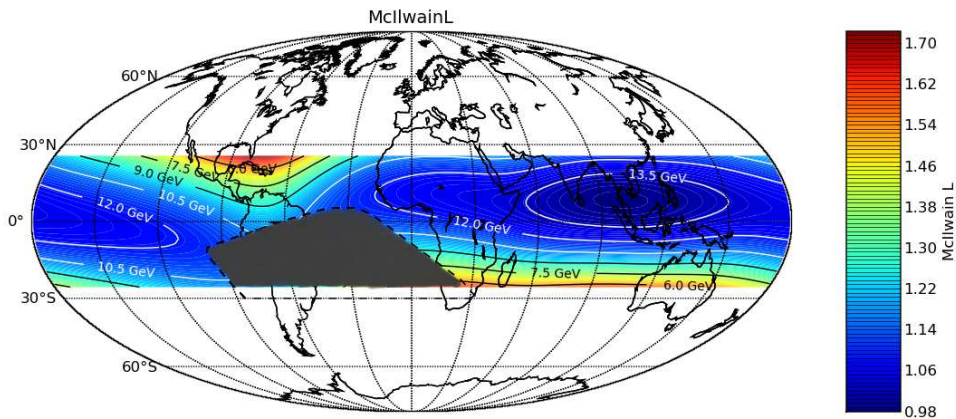}
\caption{Map of McIlwain L values for the Fermi orbit. Overlaid in 
    contours are the corresponding values for vertical rigidity cutoff. These
    values were calculated using the $10^{th}$ generation IGRF model. It
    is important to note that the model is not valid inside the South Atlantic 
    Anomaly (SAA) region, illustrated by the dark gray polygon in this 
    figure.}
\label{fig:McIlwainLDistribution}\label{GeomagneticEnv}
\end{figure*}
In the routine data taking configuration the onboard filter instance
providing the main data source for photon science (the \filter{GAMMA})
is configured to pass all the events depositing more than 20 GeV in the
calorimeter. Therefore it constitutes the main source of data for the
high-energy electron analysis, as well.
In addition to that, an unbiased sample of all trigger types, prescaled by
a factor of 250, is also downlinked for diagnostic purposes 
(through the so called \filter{DGN}); it is perfectly
adequate to study the electrons at relatively low energies, where the
higher fluxes compensate for the lower statistics due to the event prescaling.
The events collected through the \filter{DGN} are essential for the extension
of the GCR electron spectrum, in fact they constitute the main source of 
electrons for this study.
\section{Geomagnetic environment}
The Earth's field is closely approximated by the field of a dipole 
positioned near the center of the Earth. For a dipole field, 
\begin{equation}\label{RTheta}
R = R_0\cos^2\theta
\end{equation}
defines lines of constant magnetic field, also known as ``L shells''. 
${\rm R_0}$ is the 
radial distance to the field line where it crosses the geomagnetic equator 
and R is the radial distance to the point where the field is B at latitude 
$\theta$. From equation \ref{RTheta} it is possible to define the coordinate
L, which is defined as 
${\rm L} = {\rm R_0}/{\rm R_E}$ where ${\rm R_E}$ is the Earths's radius 
(6371 km)~\cite{WaltBook}. Therefore, positions around the Earth with the same
value of L are magnetically equivalent. Due to the fact that the 
Earth's dipole is offset and tilted with respect to the center of the Earth,
L values must be calculated based on a detailed field model~\cite{IGRF} 
and are a function of geomagnetic latitude~\cite{WaltBook}. The shielding 
effect of the Earth's magnetic field is smaller (larger) for larger (smaller) 
values of L and with Fermi's nearly 
circular orbit it is possible to measure the GCR electron spectrum in the range 
1.0 $<$ L $<$ 1.73 which translates into a minimum rigidity of $\sim$ 6 GeV.
Figure \ref{fig:McIlwainLDistribution} illustrates the distribution of 
L and the corresponding vertical rigidities for the Fermi orbit,
based on the IGRF model~\cite{IGRF}. It important to stress here that 
the vertical cutoff rigidity distribution shown in 
figure \ref{fig:McIlwainLDistribution} is simply an illustration and
we do not depend on this model in any way for this analysis.  

Fermi's orbit intersects the South Atlantic Anomaly (SAA) which is a region
of the Earth's radiation belt which features geomagnetically trapped
protons with energies up to hundreds of MeV and electrons with energies
up to tens of MeV. The extreme conditions within this region
impose constraints on the LAT operations, in particular the triggering, 
recording and transmission of science data are stopped during the SAA passages
~\cite{Eduardo}. The SAA is represented by the dark shaded region in figure
\ref{fig:McIlwainLDistribution}.

\section{Event selection}
The LAT is sensitive to electrons over more than five orders
of magnitude in energy ($\sim 100$ MeV -- $\sim 1$ TeV).
Across this huge energy range both the typical event topology and the relative
fractions of signal and background in the cosmic-ray flux impinging on the
detector undergo dramatic variations.
For these reasons it's not trivial to develop a single, unitary analysis
strategy providing the necessary electron detection efficiency and the hadron
rejection power across the whole energy range.
In fact we have elaborated two independent electron selections,
tuned for relatively low energies and high energies respectively (which
we shall refer to as \lowe\ and \highe\ in the rest of the paper).
The split point, in energy, between these two analyses can be naturally placed
at 20 GeV for at least two reasons.
The first one is that the onboard filtering is essentially disengaged above
this energy (whereas we only have a prescaled unbiased sample below 20 GeV).
The other one is that below $\approx 20$ GeV,
the shielding effect of the geomagnetic field (which depends on the
geomagnetic latitude and, hence, on the position of the spacecraft across its
orbit) becomes important. In fact the \highe\ selection has been \emph{mostly}
developed exploiting the data source provided by the \filter{GAMMA} for the
purpose of measuring the GCR electrons, whereas the \lowe\
selection was designed to measure both the GCR and the electron albedo 
population, using the \filter{DGN} data source. 

It must be stressed, however, that these divisions
are partially \emph{artificial} and there's a significant overlap in energy
between the two selections. This can be used as an independent cross check 
for both of the analyses. 
\begin{figure}
\centering
\includegraphics[width=75mm]{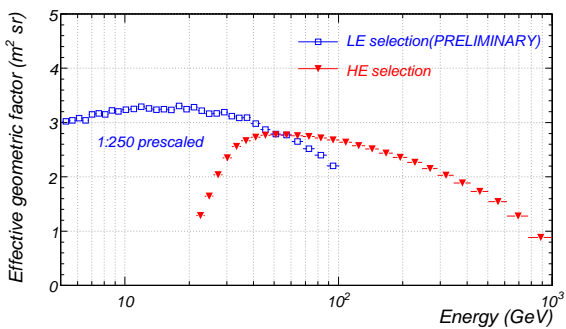}
\caption{Effective Geometric Factor for \lowe\ (blue squares, multiplied 
  by a factor 250)  and \highe\ events (red triangles). The results for
  the \lowe\ selection are still preliminary.}
\label{fig:GeometryFactor}
\end{figure}
The details of the selection criteria for the \lowe\ selection can be found
in~\cite{pescerollins-2009} and for the \highe\ in ~\cite{PRL}. 

One of the important
figures of merit which come out of the selection is the effective geometry
factor, defined as the effective area integrated over the field of view of the
detector, and is shown in figure \ref{fig:GeometryFactor}. The \lowe\ geometry 
factor has multiplied by the \filter{DGN} prescale factor of 250 for 
graphical clarity 
and in order to give an idea of the relative electron efficiencies of the 
two selections. The shape of the geometry factor is almost entirely due
to the event selection that must obtain a good background rejection power
in the whole energy range.
The sharp decrease in \highe\ selection just below 30 GeV is mainly due
to the \filter{GAMMA} onboard event filtering designed to remove 
charged particles 
with deposited energy lower that 20 GeV. 
 
\section{Spectrum reconstruction}
The electron flux is evaluated by subtracting the
residual background from the count rate and then correcting the
observed number of events with the geometry factor.
\begin{figure*}[t]
\centering
\includegraphics[width=78mm]{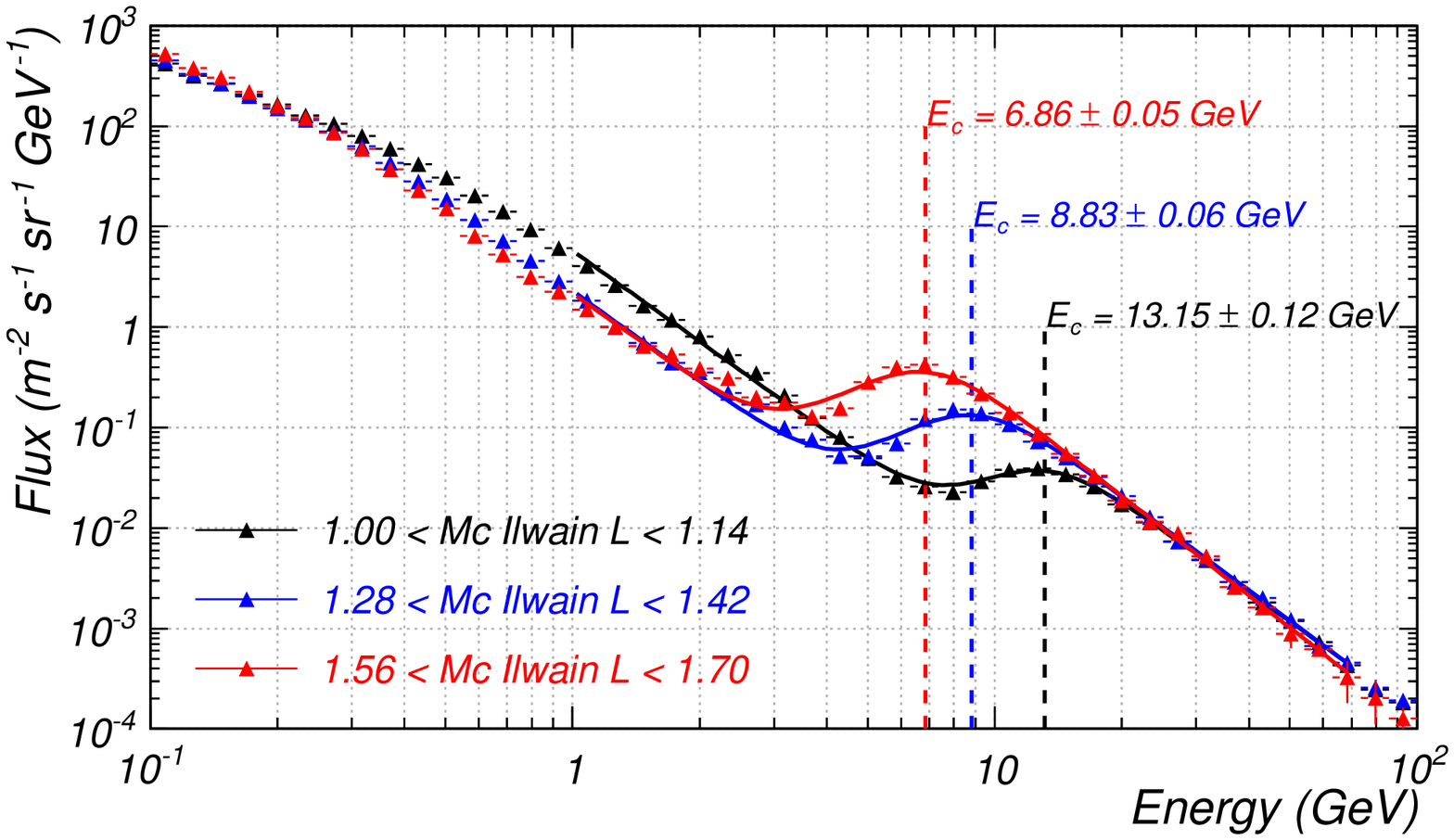}
\includegraphics[width=78mm]{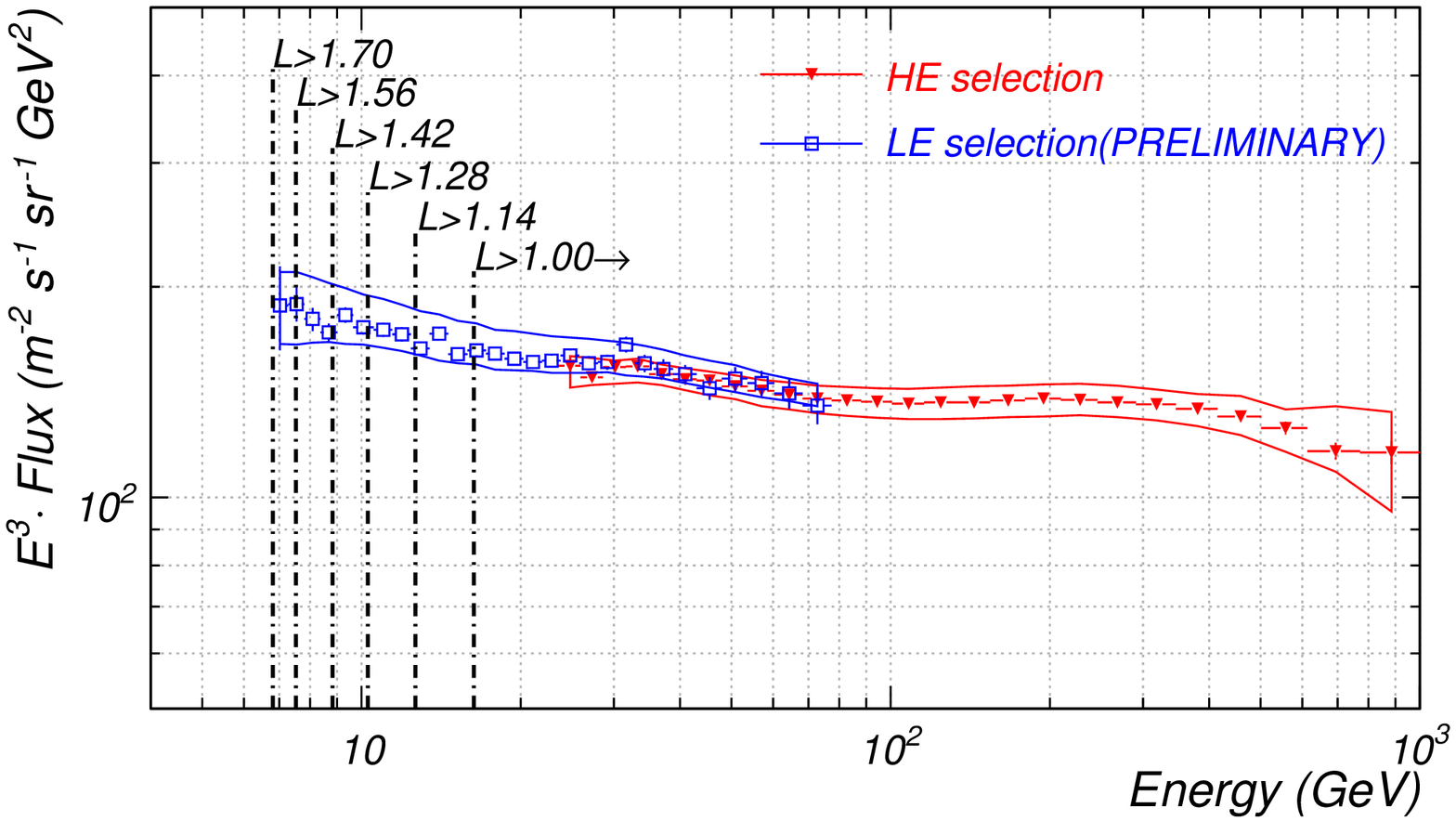}
\caption{Left panel: The measured electron flux in three McIlwain L bins. 
    For each bin the fit of the flux with equation \ref{eq:spectrum_function} 
    and the resulting estimated cutoff rigidity, E$_c$, is shown. As 
    described in the text, E$_c$ decreases for larger values of McIlwain L.
    Right panel: Illustration of the reconstruction technique used to 
    measure the 
    the primary cosmic-ray electron spectrum down the lowest accessible energy
    for the Fermi orbit. The blue points represent the GCRE flux multiplied
    by the energy cubed obtained via the \lowe\ analysis and the red points
    are from the \highe\ analysis. The solid lines depict the systematic 
    uncertainties and the vertical dashed lines serve to show from which
    McIlwain L region of the orbit the flux was measured. These results are
    preliminary.}
\label{fig:McIlwainLSpectra}
\end{figure*}
The evaluation of the residual hadron background takes
advantage of the detailed on-orbit 
particle environment model built by the LAT collaboration.
This highly detailed model has been intensively used to develop all the
$\gamma$-ray background rejection algorithms, both on-orbit and off-line.
The model includes cosmic rays and earth albedo $\gamma$-rays 
starting from 10 MeV and is valid outside
the South Atlantic Anomaly (since the LAT does not take data while passing 
through the SAA).
The particle fluxes are chosen to fit experimental data 
of several past experiments, more details can be found in~\cite{LAT}.

The event selection is tuned to keep the residual contamination below 
about 20\% in the whole energy range.
The sample of events passing the \filter{GAMMA}  are the main
source of GCR electrons given that these are all those events with
raw energy $> 20$ GeV and for these energies the effects of the Earth's
magnetic field on the incoming GCR electron flux are negligible. In order to
extend the energy range of the measured spectrum (down to lower energies)
it is necessary to use the events passing the \filter{DGN}. This extension 
is not a straight forward task and requires to take advantage of the
fact that the rigidity cutoff varies as a function of orbital
position and consequently measure the spectra in various bins of geomagnetic
position. 
As already mentioned in section~\ref{GeomagneticEnv}, the shielding effect 
of the Earth's magnetic field is smaller for larger values of the L 
coordinate and therefore to 
measure electrons of Galactic origin with energies in the lower end of our
rigidity cutoff range it is necessary to sample all those events 
collected when the detector was located at large values of L. 
Each L bin has an associated rigidity cutoff value and it is
possible to measure this value by considering that the shape of the primary 
spectrum around the geomagnetic rigidity cutoff can be parametrized as:
\begin{equation}
f_{\rm c}(E) \simeq \frac{1}{1 + (E/E_{\rm c})^{-6}}
\end{equation}
Therefore the full spectrum can be fitted with a function of the form:
\begin{equation}\label{eq:spectrum_function}
\frac{{\rm d}N}{{\rm d}E} = c_{\rm s}E^{-\Gamma_{\rm s}} +
\frac{c_{\rm p}E^{-\Gamma_{\rm p}}}{1 + (E/E_c)^{-6}}
\end{equation}
Where the subscript $s$ stands for \emph{secondary} and represents the
albedo population of electrons while the subscript $p$ stands for \emph{primary}
and represents the Galactic component of the spectra and $\Gamma$ is the
spectral index. The value of E$_{c}$, obtained by the fit is an estimation 
of the rigidity cutoff. An example of such a fit is shown in 
figure \ref{fig:McIlwainLSpectra} for three L bins.
Due to the complexity of particle orbits in the Earth's magnetosphere, the 
geomagnetic cut-off is not sharp but rather a smooth transition and therefore
it is necessary to include a multiplicative factor, $X$, to the value of 
E$_c$ obtained by the fit. This factor serves as a sort of 
\emph{padding}, and its value is driven by the need to balance 
between two requirements, namely to measure the lowest possible Galactic 
CRE energy while at the same time making sure that the measured flux value
is not affected by the shielding due to the Earth's magnetic field. 
The value of $X$ which satisfies these requirements was chosen by performing 
a scan on several values of $X$ around unity. Having done this we are able to
quantify the changes in the flux level induced by the value of $X$ used to 
extend the spectrum. From this study we found that a value of 1.15 satisfies
the above requirements. 

Once the value for $E_c$ has been estimated for each L bin and
the \emph{padding} size has been fixed, it is possible to extend the GCRE
spectrum. An example of how this is done is shown in the
right panel of figure \ref{fig:McIlwainLSpectra} where the vertical dashed 
lines denote the L slice from which the flux was measured. Where L$>$1.00
corresponds to the orbital averaged data (i.e. where the shielding effects from
the Earth's magnetic field are negligible) and L$>$1.70 is the highest value 
for L sampled by the Fermi orbit. The spectra obtained with \lowe\ and the 
\highe\ are shown in this figure to illustrate the very good agreement in
the overlap region. The solid (blue and red) lines represent the systematic 
uncertainties for the two selections.

\section{Discussion}
\begin{figure*}[t]
\centering
\includegraphics[width=132mm]{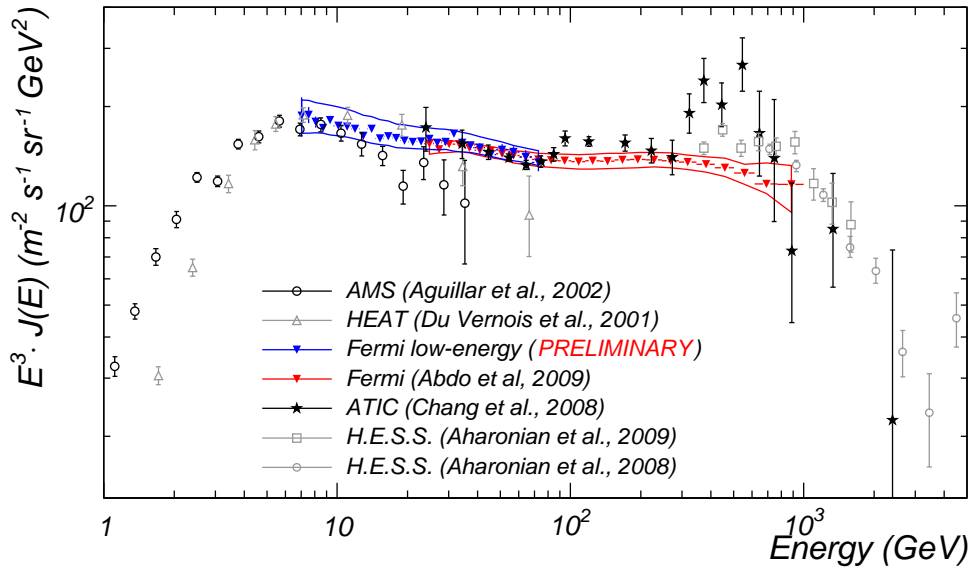}
\caption{The GCR electron spectrum measured by the Fermi LAT, red points
  correspond to the \highe\ selection and blue points to the \lowe. For
  comparison the measurements from the experiments AMS, HEAT, ATIC and 
  HESS are also shown. The extension of the Fermi LAT measurement is
  in good agreement within the systematic errors with the measurements
  made by AMS and HEAT.}
\label{fig:ExtendedSpectrum}
\end{figure*}

The GCR electron spectrum measured by the Fermi LAT from $\sim$ 7 GeV to
1 TeV is shown in figure \ref{fig:ExtendedSpectrum} together with the 
measurements made from ground, balloon and spaced based experiments. The
overlap region between the \lowe\ and the \highe\ selections for energies 
between 20 and 80 GeV shows very good agreement and serves as an 
excellent cross check between the two separate analyses. The low energy part of 
the spectrum measured by Fermi shows the same rising trend in flux as 
both the AMS and HEAT measurements and is compatible with those measurements
within the systematic errors. The extension of the GCR electron spectrum
provides valuable information needed to make constraints on possible 
propagation models of the GCRs. However, a discussion on the interpretation
of this new measurement is beyond the scope of this paper.   

\bigskip 
\begin{acknowledgments}
The $Fermi$ LAT Collaboration acknowledges generous ongoing support 
from a number of agencies and institutes that have supported both 
the development and the operation of the LAT as well as scientific data 
analysis.  
These include the National Aeronautics and Space Administration and 
the Department of Energy in the United States, 
the Commissariat \`a l'Energie Atomique and 
the Centre National de la Recherche 
Scientifique / Institut National de Physique Nucl\'eaire et 
de Physique des Particules in France, the Agenzia Spaziale Italiana 
and the Istituto Nazionale di Fisica Nucleare in Italy, 
the Ministry of Education, Culture, Sports, Science and Technology (MEXT), 
High Energy Accelerator Research Organization (KEK) and 
Japan Aerospace Exploration Agency (JAXA) in Japan, and 
the K.~A.~Wallenberg Foundation, the Swedish Research Council 
and the Swedish National Space Board in Sweden.

Additional support for science analysis during the operations phase 
is gratefully acknowledged from the Istituto Nazionale di Astrofisica in Italy.
\end{acknowledgments}

\bigskip 
\bibliography{biblio_fermicre}
\bibliographystyle{plain}

\end{document}